Larry Wos*

Dolph Ulrich

Branden Fitelson *

# XCB, the Last of the Shortest Single Axioms for the Classical Equivalential Calculus

**Abstract**. It has long been an open question whether the formula XCB = EpEEEpqErqr is, with the rules of substitution and detachment, a single axiom for the classical equivalential calculus. This paper answers that question affirmatively, thus completing a search for all such eleven-symbol single axioms that began seventy years ago.

## 1. The Question and its History

In 1933, Lukasiewicz discovered the first formulas of length eleven capable of serving, with the rules of substitution and detachment (from $E\alpha\beta$ and $\alpha$, infer $\beta$), as single axioms for the classical equivalential calculus. They are *EEpqEErqEpr*, *EEpqEEprErq*, and *EEpqEErpEqr*. He presented his proofs in [Lukasiewicz1939] and showed as well that no shorter formulas can serve as single axioms. Three decades later, Meredith [Meredith1963] published seven additional single axioms of the same length: *EEEpqrEqErp*, *EpEEqEprErq*, *EEpEqrErEpq*, *EEpqErEEqrp*, *EEpqErEEqrp*, *EEEpEqrrEqp*, and *EEEpEqrqErp*.

In the mid-1970s, Kalman and his student Peterson undertook a computer-assisted investigation of all 630 eleven-symbol equivalential theses (distinct up to alphabetical variance). Kalman [Kalman1978] discovered[1] one additional single axiom among these, *EpEEqErpErq*, while Peterson [Peterson1977] was able to eliminate from consideration 612 of those that remained. He thus posed as open questions the status of just seven formulas of length eleven.

Six of Peterson's seven open questions were answered shortly thereafter. Wos, Winker, et al. [Wos1983] showed that four of Peterson's seven formulas are too weak and reported Winker's result that two of the remaining three, *EpEEqrEEprq* and *EpEEqrEErpq*, are shortest single axioms. The one formula whose status has remained unknown[2] since 1977 is

$$\text{XCB} = \textit{EpEEEpqErqr}.$$

---

*Work supported by the Mathematical, Information, and Computational Sciences Division subprogram of the Office of Advanced Scientific Computing Research, U.S. Department of Energy, under Contract W-31-109-Eng-38.

[1] Kalman himself graciously writes that this "evidently corrects a misprint in" [Meredith1963].

A different but related question concerning this same problematic formula arises in Hodgson's work [Hodgson1996]. Also studying with Kalman, Hodgson investigated each of the eleven-character equivalential theses in the presence of substitution, detachment, and the additional rule of reverse detachment (from $E\alpha\beta$ and $\beta$, infer $\alpha$). Despite the large number of positive results Hodgson discovered, XCB was again recalcitrant and its status as a single equivalential axiom, even with this extra rule available, was offered in [Hodgson1996] as another open question.

In the next section, we answer Peterson's question affirmatively with a proof showing that, from XCB, the pair of formulas *EEpqEEqrEpr* and *EEpqEqp* follows. This pair is known [Wos1990] to provide a complete axiomatization for classical equivalence. As it happens, an affirmative answer to Hodgson's question will emerge along the way.

Our first proof (answering both questions) was found with the aid of the automated reasoning program OTTER [McCune1994][3] and then refined substantially. In a companion paper [Wos2002], we discuss details concerning the discovery of the original proof, including methodology and strategy (cf. [Wos1999]) as well as its refinement.

## 2. The Proof

Following [Meredith1963], [Kalman1978], [Wos1983], and [Hodgson1996], we employ C. A. Meredith's rule of condensed detachment [Meredith1963], which combines detachment with a certain amount of substitution, and write *Dm.n* for the most general result of detaching formula *n* (or a substitution instance of it) as minor premiss from formula *m* (or a substitution instance of it) as major premiss. This rule not only permits succinct presentations of proofs but also is ideally suited to computer implementation. Of course its use in place of the familiar rules of substitution and detachment is wholly justified since it is well known (see, e.g., [Kalman1983]) that every formula deducible from any set of axioms by the rules of substitution and detachment is a substitution instance of a formula derivable from that set by condensed detachment alone.

---

[2] The claim in [Wos1983] that XCB is too weak to be a single axiom is corrected in [Wos1999]. A fuller discussion in [Wos2002] explains why the authors of [Wos1983] were led to believe otherwise.

[3] That such assistance was invaluable, and perhaps indispensable, will occur to the reader who attempts to carry out by hand the condensed detachment of line 16 from line 12 to obtain line 17 in the proof given in the following section. The substitution instances of 12 and of 16 required for that condensed detachment are, respectively, 2,939 and 2,919 symbols in length, a consideration that may explain in part why these two questions about XCB remained unanswered for so long.

To aid readability of the longer theorems involved in the proof, when a derived theorem contains one or more alphabetical variants of XCB as a subformula, we replace the first such subformula in that theorem with an occurrence of the letter "**A**", the second with "**B**", and so forth. In these cases, each upper-case letter replaces a variant of XCB that contains only variables not occurring elsewhere in the theorem in question.

|  |  | 1. *EpEEEpqErqr* |
|---|---|---|
| D1.1 | = | 2. *EEE**A**sEtst* |
| D2.1 | = | 3. *EEE**A**stEst* |
| D1.3 | = | 4. *EEEEEE**A**stEstuEvuv* |
| D3.1 | = | 5. *EpEEEE**A**ptEutu* |
| D4.1 | = | 6. *EEEEEE**A**stEstuvEuv* |
| D3.4 | = | 7. *EpEEEE**A****B**pwExwx* |
| D1.5 | = | 8. *EEEEpEEEE**A**ptEutuvEwvw* |
| D2.7 | = | 9. *EEE**A****B**E**C**yzEyz* |
| D6.8 | = 10. *EEpEqEEEE**A****B**qxEyxyp* |
| D4.9 | = 11. *EEEE**A****B**vwEvw* |
| D1.10 | = 12. *EEEEEpEqEEEE**A****E****A**qxEyxypzEz$_1$zz$_1$* |
| D11.3 | = 13. *EpE**A**p* |
| D1.13 | = 14. *EEEEpE**A**ptEutu* |
| D9.14 | = 15. *EEEEEpE**A**ptuEtu**B*** |
| D9.15 | = 16. *EEEEEEpE**A**p**B**wxEwx**C*** |
| D12.16 | = 17. *EE**A****E**pE**A**tEpt* |
| D17.2 | = 18. *EE**A**EEEstEutus* |
| D3.18 | = 19. *EEEEpqErqrp* |

We interrupt the proof at this point to note, in passing, that formula 19 is shown in [Hodgson1996] (where it is called "CXM") to be a single axiom for classical equivalence when both condensed detachment and reverse condensed detachment are present.[4] Since it has here been obtained by condensed

---
[4] As a point of possible interest, we remark that lines 5, 7, 10, and 17 of this proof are themselves single axioms for classical equivalence when condensed detachment alone is used. If reverse condensed detachment is available as well, then so also are lines 2-4, 6, 8, and 11-14.

detachment alone, XCB is therefore also such an axiom, which answers Hodgson's question.

We now conclude our proof, thus answering Peterson's question.[5]

D1.19 =   20. *EEEEEEpqErqrpsEtst*
D20.5 =   21. *EEEAEEEEEstEutusvwEvw*
D20.19 = 22. *EEEpqpq*
D21.19 = 23. ***EEpqEEqrEpr***
D23.22 = 24. *EEpqEEErprq*
D23.24 = 25. *EEEEEpqprsEEqrs*
D25.19 = 26. ***EEpqEqp***

Since 23 and 26 together axiomatize classical equivalence, the proof that XCB is the fourteenth, and final, shortest single axiom for the equivalential calculus is complete.

**References**

[Hodgson1996] HODGSON, K., 'Shortest single axioms for the equivalential calculus with CD and RCD', *Journal of Automated Reasoning* 20 (1996), 283-316.

[Kalman1978] KALMAN, J. A., 'A shortest single axiom for the classical equivalential calculus', *Notre Dame Journal of Formal Logic* 19 (1978), 141-144.

[Kalman1983] KALMAN, J. A., 'Condensed detachment as a rule of inference', *Studia Logica* 42 (1983), 443-451.

[Lukasiewicz1939] LUKASIEWICZ, J., 'Der Äequivalenzenkalkül', *Collectanea Logica* 1 (1939), 145-169. English translation in S. MCCALL (ed.), *Polish Logic, 1920-1939*, Clarendon Press, Oxford,1967, pp. 88-115; also in L. BOROKOWSKI (ed.), *Jan Lukasiewicz: Selected Works*, North-Holland, Amsterdam, 1970, pp. 256-277.

[McCune1994] MCCUNE, W., *OTTER 3.0 Reference Manual and Guide*, Technical Report ANL-94/6, Argonne National Laboratory, Argonne, Illinois, 1994.

---

[5] Once line 19 has been reached, Peterson's question can also be answered by showing that the rule of reverse detachment is derivable when XCB is taken as the sole axiom and substitution and detachment are the rules of inference. In fact, we obtained the latter result first. Assume that $E\alpha\beta$ and $\beta$ are both theorems. From the appropriate substitution instance of XCB, the detachment of $\beta$ provides $EEE\beta\beta E\alpha\beta\alpha$. It is straightforward to prove that $Epp$ is also a theorem. Detaching $Epp$ from the requisite instance of XCB gives $EEEEppqErqr$, and from suitable instances of that and of XCB itself, $EEEEppqrEqr$ follows. From this, and the theorem $EEE\beta\beta E\alpha\beta\alpha$ obtained earlier, $EE\alpha\beta\alpha$ follows. But $E\alpha\beta$ is provable, by hypothesis, whence detachment delivers $\alpha$, completing the proof.

[Meredith1963] MEREDITH, C. A., and A. N. PRIOR, 'Notes on the axiomatics of the propositional calculus', *Notre Dame Journal of Formal Logic* 4 (1963), 171-187.

[Peterson1977] PETERSON, J. G., *The Possible Shortest Single Axioms for EC-tautologies*, Report 105, Department of Mathematics, University of Auckland, 1977.

[Wos1983] WOS, L., S. WINKER, R. VEROFF, B. SMITH, and L. HENSCHEN, 'Questions concerning possible shortest single axioms for the equivalential calculus: An application of automated theorem proving to infinite domains', *Notre Dame Journal of Formal Logic* 24 (1983), 205-223.

[Wos1990] WOS, L., 'Meeting the challenge of fifty years of logic', *Journal of Automated Reasoning* 6 (1990), 213-232.

[Wos1999] WOS, L., with G. PIEPER, *A Fascinating Country in the World of Computing: Your Guide to Automated Reasoning*, World Scientific, Singapore, 1999.

[Wos2002] WOS, L., D. ULRICH, and B. FITELSON, 'An open question for XCB in equivalential calculus answered', Preprint ANL/MCS-P964-0602, Mathematics and Computer Science Division, Argonne National Laboratory, Argonne, Illinois, June 2002 (under review by the *Journal of Automated Reasoning*).

LARRY WOS
Mathematics and Computer Science Division
Argonne National Laboratory
Argonne, IL  60439, U.S.A.
wos@mcs.anl.gov

DOLPH ULRICH
Department of Philosophy
Purdue University
West Lafayette, IN  47907-1360, U.S.A.
dulrich@purdue.edu

BRANDEN FITELSON
Philosophy Department
San Jose State University
San Jose, CA  95192, U.S.A.
branden@fitelson.org